\documentstyle[aps,graphicx,multicol]{revtex}

\draft

\begin{document}

\title{Entropy in hot $^{161,162}$Dy and $^{171,172}$Yb nuclei}

\author{M.~Guttormsen, A.~Bjerve, M.~Hjorth-Jensen, E.~Melby, J.~Rekstad, A.~Schiller and S.~Siem}

\address{Department of Physics, University of Oslo, N-0316 Oslo, Norway}

\author{A.~Belic}

\address{Institute of Physics, YU-11001 Belgrade, Yugoslavia}

\maketitle

\begin{abstract}

The density of accessible levels at low spin in the ($^3$He,$\alpha \gamma$) reaction has been extracted for the $^{161,162}$Dy and $^{171,172}$Yb nuclei. The entropy of the even-odd and even-even nuclei has been deduced as a function of excitation energy, and found to reach a maximum of 15 $k_B$ before neutron evaporation. The entropy of one quasi-particle outside an even-even core is found to be 1.70(15) $k_B$. This quasi-particle picture of hot nuclei is well accounted for within a simple pairing model. The onset of two, four and six quasi-particle excitations in the $^{162}$Dy and $^{172}$Yb nuclei is discussed and compared to theory. The number of quasi-particles excited per excitation energy is a measure for the ratio of the level energy spacing and the pairing strength. 

\end{abstract}


\begin{multicols}{2}

\section{Introduction}

Already in 1936 H.~Bethe introduced the Fermi gas to describe nuclear properties at high temperature \cite{1}. This simple independent particle model has been modified by including residual interactions between the nucleons. In the low excitation region long-range pair correlations play an important role and are roughly described within the so-called back-shifted Fermi gas model \cite{2}.

There is evidence for the existence of paired nucleons (Cooper pairs) at low temperature. In high spin physics, the backbending phenomenon is a beautiful manifestation of the breaking of pairs. The mechanism is induced by Coriolis forces tending to align single particle angular momenta along the nuclear rotational axis \cite{3,4}. Theoretical models also predict reduction in the pair correlations at higher temperatures \cite{5,6,7}.

The breaking of pairs is difficult to observe as a function of intrinsic excitation energy. Recent theoretical \cite{7} and experimental \cite{8,9} works indicate that the process of breaking pairs takes place over several MeV of excitation energy. The corresponding critical temperature is measured to be $T_c \sim $ 0.5 MeV/$k_B$ \cite{10}, where $k_B$ is Boltzmann's constant.

The aim here is to extract the entropy of the $^{161,162}$Dy and $^{171,172}$Yb isotopes, and deduce the number of excited quasi-particles as function of excitation energy. In Sect.~2 we describe the experimental techniques and analyzing tools. Sect.~3 presents results for the entropy using a simple pairing Hamiltonian with an even and odd number of fermions distributed over $L$ single-particle levels with double degeneracy. This is a model, which for small numbers of fermions, typically less than 20, can be solved numerically yielding all possible eigenstates. It results in the full level density and can in turn be used to extract thermodynamical quantities. Since we expect pairing correlations to be important in nuclei, such a simple model should mimic to a certain extent the entropy extracted from the experimental level density. In Sect.~4 we present the experimental findings and relate them to the simple pairing model of Sect.~3. Concluding remarks are given in Sect.~5.

\section{Experimental methods}

The Oslo cyclotron group has developed a method to extract nuclear level densities at low spin from measured $\gamma$-ray spectra \cite{8,9,10,11,17}. The main advantage of utilizing $\gamma$-rays as a probe for level density is that the nuclear system is likely thermalized prior to the $\gamma$-emission. In addition, the method allows for the simultaneous extraction of level density and $\gamma$-strength function over a wide energy region.

The experiments were carried out with 45~MeV $^3$He-projectiles accelerated by the MC-35 cyclotron at the University of Oslo. The experimental data were recorded with the CACTUS multidetector array \cite{12} using the ($^3$He,$\alpha \gamma$) reaction on $^{162,163}$Dy and $^{172,173}$Yb self-supporting targets. The charged ejectiles were detected with eight particle telescopes placed at an angle of 45$^{\circ}$ relative to the beam direction. Each telescope comprises one Si~$\Delta E$ front and one Si(Li) $E$ back detector with thicknesses of 140 and 3000~$\mu$m, respectively. 

An array of 28 NaI $\gamma$-ray detectors with a total efficiency of $\sim$15\% surrounded the target and particle detectors. In addition, two Ge detectors were used to monitor the spin distribution and selectivity of the reactions.

From the reaction kinematics the measured $\alpha$-particle energy can be transformed to excitation energy $E$. Thus, each coincident $\gamma$-ray can be assigned to a $\gamma$-cascade originating from a specific excitation energy. The data are sorted into a matrix of $(E,E_{\gamma})$ energy pairs. At each excitation energy $E$ the NaI $\gamma$-ray spectra are unfolded \cite{13}, and this matrix is used to extract the first-generation (or primary) $\gamma$-ray matrix with the well-established subtraction technique of Ref.~\cite{14}. 

The resulting matrix $P(E,E_{\gamma})$, which describes the primary $\gamma$-spectra obtained at initial excitation energy $E$, is factorized according to the Brink-Axel hypothesis \cite{15,16} by 
\begin{equation}
P(E,E_{\gamma}) \propto \rho (E -E_{\gamma})\sigma (E_{\gamma}) . 
\end{equation}
The assumptions and methods behind the factorization of this expression are described in Refs.~\cite{11,17}, and only a short outline is given here.

Both the level density $\rho$ and the $\gamma$-energy dependent function $\sigma$ are unknown. In the new iteration procedure \cite{17} the first $\rho^0$ function is simply taken as a straight line, and the first $\sigma^0$ is calculated from Eq.~(1). Then new $\rho$ and $\sigma$ functions are analytically calculated by minimizing the least square fit $\chi ^2$ to the data set $P$. This procedure is repeated until a global minimum is obtained with respect to the values at all ($E$, $E_{\gamma}$) pairs. About 50 iterations are necessary for fitting the $\sim$ 150 free parameters to the $\sim$ 1500 data points of $P$. Due to methodical problems in the first-generation procedure, we only use data with $\gamma$-energies $E_{\gamma}> 1$ MeV and excitation energies $E >$ 2.5 and 4.0 MeV in the odd-even and even-even isotopes, respectively \cite{17}.

It has been shown \cite{17} that if one solution for $\rho$ and $\sigma$ is known, it is possible to construct infinitely many solutions with the same $\chi ^2$ using the substitution
\begin{equation}
\rho (E -E_{\gamma}) \rightarrow A \exp [\alpha(E-E_{\gamma})] \rho (E -E_{\gamma})
\end{equation}
and
\begin{equation}
\sigma (E_{\gamma}) \rightarrow B\exp (\alpha E_{\gamma})\sigma (E_{\gamma}),
\end{equation}
where $A$, $B$ and $\alpha$ are arbitrary parameters. In the new product of these two functions, a factor $AB\exp (\alpha E)$ is left over, which is absorbed in $P$, since the sum $\sum_{E_{\gamma}} P(E,E_{\gamma})$ is undetermined.

In the case of $^{162}$Dy, Fig.~1 demonstrates how the parameters $A$ and $\alpha $ are determined to obtain a level density function (data points) with correct number of levels around the ground state (histogram). In addition, the parameters reproduce the level density calculated from the spacing of neutron resonances at the neutron binding 
energy $B_n$, see insert of Fig.~1.

In the following, we concentrate only on the information given by the level density, which is assumed to be independent of particular $\gamma$-ray decay routes.

\section{Entropy from a simple pairing model}

The level density\footnote{Hereafter we use $\rho$ for the level density in the microcanonical ensemble. Furthermore, since we are dealing with a system with discrete 
energies from a quantal system, the microcanonical partition function is
defined by the number of states at a given energy $E$ \cite{gross99}.} $\rho$ defines the partition function for the microcanonical ensemble, the latter being the appropriate one for statistical descriptions of isolated systems such as finite nuclei. The partition function for the canonical ensemble is related to that of the microcanonical ensemble through a Laplace transform
\begin{equation}
     Z(\beta)=\int_0^{\infty}dE\rho(E)\exp{(-\beta E)}.
     \label{eq:zcan}
\end{equation}
Here we have defined $\beta=1/k_BT$, where $T$ is the temperature and $k_B$ is Boltzmann's constant. Since we will deal with discrete energies, the Laplace transform of Eq.\ (\ref{eq:zcan}) takes the form
\begin{equation}
         Z(\beta)=\sum_E \Delta E\rho(E)\exp{(-\beta E)},
         \label{eq:zactual}
\end{equation}
where $\Delta E$ is the energy bin used.

In nuclear and solid state physics, thermal properties have mainly been studied in the canonical and grand-canonical ensemble. In order to obtain the level density, the inverse transformation 
\begin{equation}
      \rho(E) =\frac{1}{2\pi i}\int_{-i\infty}^{i\infty}
 d\beta Z(\beta) \exp{(\beta E)},
      \label{eq:zbigcan}
\end{equation}
is normally used. Compared with Eq.\ (\ref{eq:zcan}), this transformation is rather tricky to perform since the integrand $\exp{\left(\beta E+ \ln Z(\beta)\right)}$ is a rapidly varying function of the integration parameter. In order to obtain the density of states, approximations like the saddle-point method, viz., an expansion of the exponent in the integrand to second order around the equilibrium point and subsequent integration, have been used widely. For the ideal Fermi gas, this gives the following density of states
\begin{equation}
      \rho_{\rm ideal}(E)=\frac{\exp{(2\sqrt{aE})}}{E\sqrt{48}},
      \label{eq:omegaideal}
\end{equation}
where $a$ in nuclear physics is a factor typically of the order $a=A/8$ with dimension MeV$^{-1}$, $A$ being the mass number of a given nucleus. 

Ideally, the experiment should provide the level density as function of excitation energy and thereby the 'full' partition function for the microcanonical ensemble. In the microcanonical ensemble we could then extract expectation values for thermodynamical quantities like temperature $T$, or the heat capacity $C$. The temperature in the microcanonical ensemble is defined as 
\begin{equation}
      \langle T\rangle=\left(\frac{dS(E)}{dE}\right)^{-1}.
      \label{eq:temp}
\end{equation}
It is a function of the excitation energy, which is the relevant variable of interest in the microcanonical ensemble. However, since the extracted level density is given only at discrete energies, the calculation of expectation values like $T$, involving derivatives of the partition function, is not reliable unless a strong smoothing over energies is performed. This case is discussed at large in Ref.\ \cite{9} and below. Another possibility\footnote{The transformation to the canonical ensemble represents also a smoothing.} is to employ the Laplace transformation of Eq.~(\ref{eq:zactual}) in order to evaluate various thermodynamical quantities in the canonical ensemble. As an example, we can evaluate the entropy in the canonical ensemble using the definition of Helmholtz free energy \begin{equation}
     F(T)= -k_B T \ln Z(T)=\langle E(T)\rangle - TS(T).
\end{equation}
Note that the temperature $T$ is now the variable of interest and the energy $E$ is given by the expectation value $\langle E\rangle$ as function of $T$. Similarly, the entropy $S$ is also a function of $T$.

In this section, we extract the exact level density from a simple theoretical model\footnote{A similar analysis within the framework of several BCS ansatz based approaches was done by D{\o}ssing {\sl et al.}~in Ref.\ \cite{7}. Whereas our approach includes all possible eigenvalues in order to determine the level density, D{\o}ssing {\sl et al.}~perform their diagonalization within a space spanned by number-projected states. The qualitative behavior of their results is however similar to that presented here.}. The Hamiltonian we use to obtain the eigenvalues and the level density  $\rho(E)$ is the simple pairing Hamiltonian 
\begin{equation}
   H=\sum_i \varepsilon_i a^{\dagger}_i a_i -\frac{1}{2} G\sum_{ij>0}
           a^{\dagger}_{i}
     a^{\dagger}_{\bar{\imath}}a_{\bar{\jmath}}a_{j},
     \label{eq:pairHamiltonian}
\end{equation}
where $a^{\dagger}$ and $a$ are fermion creation and annihilation operators, respectively. The indices $i$ and $j$ run over the number of levels $L$, and the label $\bar{\imath}$ stands for a time-reversed state. The parameter $G$ is the strength of the pairing force while $\varepsilon_i$ is the single-particle energy of level $i$. 

We assume that the single-particle levels are equidistant with a fixed spacing $d$ such that Eq.\ (\ref{eq:pairHamiltonian}) becomes
\begin{equation}
   H=d\sum_{i} ia^{\dagger}_i a_i 
     -\frac{1}{2}G\sum_{ij>0}a^{\dagger}_i
     a^{\dagger}_{\bar{\imath}}a_{\bar{\jmath}}a_{j}.
     \label{eq:pair1}
\end{equation}
Moreover, in our simple model, the degeneracy of the single-particle levels is set to $2J+1=2$, with $J=1/2$ being the spin of the particle. 

Introducing the pair-creation operator 
\begin{equation}
   S^+_i=a^{\dagger}_{im}a^{\dagger}_{i-m},
\end{equation}
and
\begin{equation}
   S^-_i=a_{i-m}a_{im},
\end{equation}
one can rewrite the Hamiltonian in Eq.\ (\ref{eq:pair1}) as
\begin{equation}
   H=d\sum_iiN_i
     -\frac{1}{2} G\sum_{ij>0}S^+_iS^-_j,
     \label{eq:pair2}
\end{equation}
where  
\begin{equation}
   N_i=a^{\dagger}_i a_i
\end{equation}
is the number operator. The latter commutes with the Hamiltonian $H$. In this model quantum numbers like seniority $\cal{S}$ are good quantum numbers, and the eigenvalue problem can be rewritten in terms of blocks with good seniority. Loosely speaking, the seniority quantum number $\cal{S}$ is equal to the number of unpaired particles, see Ref.\ \cite{talmi93} for further details. 

The reason why we focus on such a simple model is twofold. Firstly, we expect the ground state of nuclei to be largely dominated by pairing correlations. This is mainly due to the strong singlet $^1S_0$ state in the nucleon-nucleon interaction, see e.g., Refs.\ \cite{pair1,pair2,pair3}. For even-even systems this is typically reflected in an energy gap between the ground state and the first excited state, a gap that is larger than that seen in odd nuclei. This is taken as an evidence of strong pairing correlations in the ground state. More energy is needed in order to excite the system when all fermions are paired, i.e. when we have a system with an even number of particles. Since pairing correlations are important in nuclear systems, we expect the above model to exhibit at least some of the properties seen in finite nuclei.

Secondly, for particle numbers up to $N \sim 20$, the above model can be solved exactly through numerical diagonalization, since seniority is a good quantum number. This means that we can subdivide the full eigenvalue problem into minor blocks with given seniority and diagonalize these separately. In our case we use for the even system $N=12$ particles which are distributed over $L=12$ two-fold degenerate levels giving a total of \begin{equation}
\left(\begin{array}{c}2L\\N\end{array}\right)=\left(\begin{array}{c}24\\12\end{array}\right)=2.704.156
\end{equation}
states. Of this total, for $\cal{S}$ $=0$, i.e. no broken pairs, we have 
\begin{equation}
\left(\begin{array}{c}L\\N/2\end{array}\right)=\left(\begin{array}{c}12\\6\end{array}\right)=924,
\end{equation} 
states. Since the Hamiltonian does not connect states with different seniority $\cal{S}$, we can diagonalize a $924\times 924$ matrix and obtain all eigenvalues with $\cal{S}$ $=0$. Similarly, we can subdivide the Hamiltonian matrix into $\cal{S}$ $=2$, $\cal{S}$ $=4$,... and $\cal{S}$ $=12$ (all pairs broken) blocks and obtain {\em all} $2.704.156$ eigenvalues for a system with $L=12$ levels and $N=12$ particles. As such, we have the {\em exact density of levels} and can compute observables like the entropy, heat capacity etc. This numerically solvable model enables us to compute exactly the entropy in the microcanonical and the canonical ensembles for systems with odd and even numbers of particles.  In addition, varying the level spacing $d$ and the pairing strength $G$, may reveal features of e.g., the entropy which are similar to those of the experimentally extracted entropy. Recall that the nuclei studied represent both even-even and even-odd nucleon systems.

Here we study two systems in order to extract differences between odd and even systems, namely by fixing the number of doubly degenerated single-particle levels to $L=12$, whereas the number of particles is set to $N=11$ and $N=12$.  Fig.\ \ref{fig:fig2} pictures the ground state of these systems and possible excited states.

These two systems result in a total of $\sim 3 \times 10^6$ eigenstates. In the calculations, we choose a single-particle level spacing of $d=0.1$, which is close to what is expected for rare earth nuclei. In this sense, if we are to assign energies with dimension MeV, our results may show some similarity with experiment. We select three values of the pairing strength, namely $G=1, 0.2$ and $0.01$, resulting in the ratio $\delta=d/G=0.1$, $\delta=d/G=0.5$ and $\delta=d/G=10$, respectively. The first case represents a strong pairing case, with almost degenerate single-particle levels. The second is an intermediate case where the level spacing is of the order of the pairing strength, while the last case results in a weak pairing case. As shown below, the results for the latter resemble to a certain extent those for an ideal gas.

The calculational procedure is rather straightforward. First we diagonalize the large Hamiltonian matrix (which is subdivided into seniority blocks) and obtain all eigenvalues $E$ for the odd and even particle case. This defines also the density of levels $\rho(E)$, the partition function and the entropy in the microcanonical ensemble. Thereafter we can obtain the partition function $Z(T)$ in the canonical ensemble through Eq.\ (\ref{eq:zactual}). The partition function $Z(T)$ enables us in turn to compute the entropy $S(T)$ using
\begin{equation}
    S(T)=k_B \ln Z(T)+\langle E(T)\rangle/T.
    \label{eq:canentropy}
\end{equation}
Since this is a model with a finite number of levels and particles, unless a certain smoothing is done, the microcanonical entropy may vary strongly from energy to energy. This is seen in Fig.\ \ref{fig:fig3}, where we plot the entropy for the odd (upper part) and even (lower part) system using $\delta = d/G= 0.5$. The entropy is given by discrete points, since we do not have eigenvalues at all energies. However, we can also perform a moderate smoothing which conserves the basic features of the model, namely an increase in entropy when pairs are broken. This was performed with a Gaussian smoothing
\begin{equation}
    \tilde{S}_i=\frac{\sum_kS_k\exp{(-(E_i-E_k)^2/2\sigma^2)}}
                {\sum_k\exp{(-(E_i-E_k)^2/2\sigma^2)}}
\end{equation}
where $S_k$ and $E_{i,k}$ are the entropies and energies from the diagonalization of the pairing Hamiltonian. $\tilde{S}$ is the smoothed entropy. With a smoothing parameter of $\sigma=0.2$ we see that the smoothed entropy still keeps track of the points where the entropy experiences an increase due to breaking of pairs.

Figure\ \ref{fig:fig3} clearly reveals the energies where two, three, four and so forth quasi-particles are created, i.e., where sudden increases in entropy take place. For the even system with the ground state at $E_{GS}=-2.44$, the first seniority ${\cal S}$ $=2$ (formation of two quasi-particles) state appears at an excitation energy of $E=2.2$, the first ${\cal S}$ $=4$ state appears at $E=4.06$ and the first ${\cal S}$ $=6$ state is at $E=5.41$. Note well that in the figures of calculations we do not show the absolute energies. If we wish to employ dimensions in MeV, the first excited state for the even system would be close to what is expected experimentally. 

For the odd system, the first excited states are just one-quasi-particle states, i.e.,
excitations of the last and least bound single-particle. Since the level spacing is much smaller around the ground state energy for the odd case (with energy $E_{GS}=-1.65$), these states appear rather close to the ground state. When a pair is broken, we create a three-quasi-particle state (one broken pair plus a quasi-particle), or seniority ${\cal S}$ $=3$ state. This appears at an excitation energy\footnote{Note that the first state with a broken pair appears at a lower excitation energy for the odd system, as expected.} of $E=2.01$, whilst the seniority ${\cal S}$ $=5$ state (two broken pairs plus one quasi-particle) appears at $E=3.58$. We note from Fig.\ \ref{fig:fig3} that at an energy of $E\sim 8-9$, the entropy starts decreasing (population inversion), reflecting thereby the limited size of our model.

For $\delta=0.5$, where the single-particle spacing is only half the pairing strength, the energy eigenvalues are fairly well distributed over the given energy range. If we decrease $\delta$ however, we approach the degenerate limit, and the eigenvalues and the entropy are sharply concentrated around those eigenvalues where pairs are broken. This is seen in Fig.\ \ref{fig:fig4} for $\delta=0.1$ for the even case with $N=12$. The odd case with $N=11$ exhibits a similar behavior. Clearly, if we wish to evaluate the temperature according to Eq.\ (\ref{eq:temp}) for $\delta=0.1$, even with a strong smoothing, we cannot obtain reliable values for e.g., $T$. Thus, rather than performing a certain smoothing, we will choose to present further results for the entropy in the canonical ensemble, using the Laplace transform of Eq.\ (\ref{eq:zactual}). 

The results for the entropy in the canonical ensemble as functions of $T$ for the above three sets of $\delta=d/G$ are shown in Fig.\ \ref{fig:fig5}. For the two cases with strong pairing, we see a clear difference in entropy between the odd and the even system. The difference in entropy between the odd and even systems can be easily understood from the fact that the lowest-lying states in the odd system involve simply the excitation of one single-particle to the first unoccupied single-particle state, and is interpreted as a single-quasi-particle state. These states are rather close in energy to the ground state and explain why the entropy for the odd system has a finite value already at low temperatures (recall also the discussion in connection with Fig.\ \ref{fig:fig3}). Higher lying excited states include also breaking of pairs and can be described as three-, five- and more-quasi-particle states. For $\delta=10$, the odd and even systems merge together already at low temperatures, indicating that pairing correlations play a negligible role. For small single-particle spacing, also the difference in energy between the first excited state and the ground state for the odd system is rather small. 

For our choice of $d$ we observe that the maximum entropy is of the order of $S\sim 14 k_B$ in the canonical ensemble, whereas in the microcanonical ensemble, see Figs.\ \ref{fig:fig3} and \ref{fig:fig4}, the maximum value is $S \sim 10-12 k_B $. Obviously, when performing the transformation to the canonical ensemble, since we have a small system, there may be larger fluctuations in expectation values like the entropy. In the limit $N\rightarrow \infty$, the two ensembles should result in equal values for $T$, $E$ and $S$,
see Ref.\ \cite{gross99} for an in depth discussion.

For $\delta=0.5$ we note that at a temperature of $k_BT \sim 0.5-0.6$, the even and odd system approach each other\footnote{If we wish to make contact with experiment, we could again assign units of MeV to $k_BT$ and $E$.}. The temperature where this occurs corresponds to an excitation energy $\langle E\rangle$ in the canonical ensemble of $\langle E\rangle \sim 4.7-5.0$. Recalling Fig.\ \ref{fig:fig3}, this corresponds to excitation energies where we have $4-6$ quasi-particles, seniority ${\cal S}$ $=4-6$, in the even system and $5-7$ quasi-particles, seniority ${\cal S}$ $=5-7$, in the odd system. The almost merging together of the even and odd systems at these temperatures, can be retraced to the features seen in Fig.\ \ref{fig:fig3}. For higher excitation energies in Fig.\ \ref{fig:fig3}, we saw that higher seniority values show less marked bumps in the entropy, indicating that the level density at higher excitation energies contains many more states and that we are getting closer to a phase where pairing plays a less significant role. 

For small systems like finite nuclei, where the size of the system is not large
compared to the range of the strong interaction, the entropy is not an extensive 
quantity, i.e., it does not scale with the size of the system \cite{gross99}.
However, if we
assume that the entropy is an extensive quantity, then $S=nS_1$, with $n$ the number of particles and $S_1$ the single-particle entropy in the canonical ensemble. In our case $S_1$ should correspond to the single-quasi-particle entropy. If we label the entropy excess $\Delta S$ as the difference between the odd and even entropies, namely $\Delta S= S_{\rm odd} - S_{\rm even}$ with $S_{\rm odd}=n_{\rm odd}S_1$ and $S_{\rm even}=n_{\rm even}S_1$, we can in turn define the  number of quasi-particles in the odd and even systems as 
\begin{equation}
      n_{\rm odd}(E) = \frac{S_{\rm odd}}{\Delta S} \ \ \ \ {\rm and} \ \ \ \ 
       n_{\rm even}(E) =\frac{S_{\rm even}}{\Delta S} \ ,
       \label{eq:quasi-particle}
\end{equation}
respectively. The odd system has one more quasi-particle 
than the even system, i.e., $n_{\rm odd} = n_{\rm even} + 1$.

Fig.\ \ref{fig:fig6} shows the number of quasi-particles in the odd and even systems for the three values of $\delta$ using the definition in Eq.\ (\ref{eq:quasi-particle}). We note that for all cases the differences between the odd and even systems remain equal and close to one, demonstrating that the entropy is an extensive quantity as function of the number of quasi-particles. Furthermore, for $\delta=0.5$ (central panel), we see that the excitation energies where $1,2,3,\dots$ quasi-particles appear, agree with the results discussed in Fig.\ \ref{fig:fig3} in the microcanonical ensemble. To give an example, for the odd system, three quasi-particle appear at an energy of $\langle E\rangle =1.8$, which should be compared with the exactly calculated one in the microcanonical ensemble of $E=2.01$. Five quasi-particles show up at $\langle E\rangle =3.4$, which again should be compared with the result obtained in the microcanonical ensemble of $E=3.58$. The agreement for the even case is slightly worse. For $\delta=0.1$, the strong pairing case, we note that more energy is needed in order to create $2,4,\dots$ and $3,5,\dots$ quasi-particles in the even and odd systems, respectively. This agrees also with the microcanonical result of Fig.\ \ref{fig:fig4}. For the weak pairing case $\delta=10$, higher seniority states appear already at low excitations energies, indicating that pairing plays a minor role, as expected.

Fig.\ \ref{fig:fig6} carries also an interesting message. If one can extract the number of quasi-particles as function of excitation energies, the steepness of the curve provides useful information about the relation between the single-particle spacing and the pairing strength. 

In summary, varying $\delta$, allows us to extract qualitative informations about thermodynamical properties such as the entropy and the number of quasi-particles in even and odd systems. Especially, two properties are worth paying attention to concerning the discussion in the next section. Firstly, for the two cases with strong pairing ($\delta=0.1$ and $\delta=0.5$), Fig.\ \ref{fig:fig5} tells us that at temperatures where we have  $4-6$ quasi-particles in the even system and $5-7$  quasi-particles in the odd system, the odd and even system tend to merge together. This reflects the fact that pairing correlations tend to be less important and we approach the non-interacting case.  For the weak pairing case, $\delta=10$, the odd and even systems yield similar results at much lower temperatures. In a simple model with just pairing interactions, it is thus easy to see where, at given temperatures and excitation energies, certain degrees of freedom prevail. For the experimental results in the next section, this may not be the case since the interaction between nucleons is much more complicated. The hope however is that pairing may dominate at low excitation energies and that the features seen in e.g., Fig.\ \ref{fig:fig5} are qualitatively similar to the experimental ones.

Secondly, we can read from Fig.\ \ref{fig:fig6} the excitation energy where different numbers of quasi-particles appear. With a realistic value for the level spacing, a comparison with experiment may tell us something about the strength of the pairing force. 

\section{Experimental results and discussion}

The experimental level density $\rho(E)$ at excitation energy $E$ is proportional to the number of levels accessible in $\gamma$-decay. For the present reactions the spin distribution is centered around $\langle J\rangle \sim$ 4.4 $\hbar$ with a standard deviation of $\sigma_J \sim$ 2.4 $\hbar$ \cite{spindis}. Hence, the entropy\footnote{The experiment reveals the level density and not the state density. Thus, also the observed entropy reveals the number of levels. The state density can be estimated by $\rho_{\rm state} \sim (2J+1)\rho_{\rm level} \sim$ 9.8 $\rho_{\rm level}$.} can be deduced within the microcanonical ensemble, using 
\begin{equation}
S(E) = k_B \ln N(E) = k_B \ln \frac{\rho(E)}{\rho_0},
\end{equation}
where $N$ is the number of levels in the energy bin at energy $E$. The normalization factor $\rho_0$ can be determined from the ground state band in the even-even nuclei, where we have $N(E) \sim$ 1 within a typical experimental energy bin of $\sim$ 0.1 MeV. 

The extracted entropies for the $^{161,162}$Dy and $^{171,172}$Yb nuclei are shown in Figs.\ \ref{fig:fig7} and \ref{fig:fig8}. In the transformation from level density to entropy we use Eq.~(21) with $\rho_0 \sim$ 3 MeV$^{-1}$. The entropy curves are rather linear, but with small oscillations or bumps superimposed. The curves terminate around 1 MeV below their respective neutron binding energies due to the experimental cut excluding $\gamma$-rays with $E_{\gamma} <$ 1 MeV. All four curves reach $S \sim$ 13 $k_B$, which by extrapolation correspond to $S \sim$ 15 $k_B$ at the neutron binding energy $B_n$. This is the maximum entropy that a nucleus in this mass region can reach before neutron emission.

The calculations for odd and even systems (see Fig.~3) show clear increases in the entropy at the excitation energies where Cooper pairs are broken. This behavior is not very pronounced in the experimental data, probably due to residual couplings in real nuclei. In particular, our pairing model excludes collective excitations, which are known to contribute strongly at low excitation energy. For $^{172}$Yb in Fig.~8 one can identify bumps at 1.5 MeV and 2.8 MeV of excitation energy, that could be interpreted as increased entropy due to the breaking of two and four quasi-particles, respectively. 

For the odd system the valence particle (or hole) is expected to perform blocking, and indeed the calculations of Fig.~3 reveal effects of smearing out the entropy structures as function of excitation energy. The smoother experimental entropy curves for $^{161}$Dy and $^{171}$Yb (see Figs.\ \ref{fig:fig7} and \ref{fig:fig8}) seems also evident, in particular for the $^{161}$Dy case. 

The experimental entropy of the even-odd system follows closely the entropy for the even-even system, but the even-odd system has an entropy excess. The difference of entropy in the even-odd system compared to the even-even system is evaluated in Fig.\ \ref{fig:fig9} for $^{161,162}$Dy and $^{171,172}$Yb. The observed entropy excesses in the 1.5 MeV $ < E < $ 5.5 MeV excitation region are $\Delta S \sim$ 1.8(1) $k_B$ and $\sim$ 1.6(1) $k_B$ for dysprosium and ytterbium, respectively. The calculations of Fig.~5 show that the entropy excesses are due to the additional degrees of freedom imposed by a valence particle (or hole). In the center panel of Fig.~5, using $\delta=0.5$, the entropy excess is $\Delta S \sim 3 k_B$ at $k_B T \sim  0 - 0.3$. However, the model is based on the density of states formed by 1/2 spin particles. The average total spin is $J\sim 1$ giving $\rho _{\rm level} \sim \rho_{\rm state}/3$. Hence, the calculated entropy excess based on level density should be on the average $\Delta S \sim (3 -\ln 3) k_B \sim 2 k_B$, which is close to experiment. The temperature region up to 0.3 MeV corresponds roughly to the experimental excitation energy region discussed, and the model thus supports that  excitation can be described by some sort of quasi-particles.

The experimental level density can be used to determine the canonical partition function $Z(T)$. However, in the evaluation of Eq.\ (\ref{eq:zactual}), we have to extrapolate the experimental $\rho$ curve to $\sim$ 40 MeV. Here, we use the back-shifted level density formula of Refs.~\cite{23,24} with
\begin{equation}
\rho=f\frac{\exp [2\sqrt{aU}]}{12\sqrt{2}a^{1/4}U^{5/4}\sigma},
\end{equation}
where the back-shifted energy is $U=E-E_1$ and the spin cut-off parameter $\sigma$ is defined through $\sigma^2 = 0.0888 A^{2/3}\sqrt{aU}$. The level density parameter $a$ and the back-shift parameter $E_1$ are defined by $a=0.21A^{0.87}$ MeV$^{-1}$ and $E_1 = C_1 + \Delta$, respectively, where the correction factor is given by $C_1=-6.6A^{-0.32}$ according to Ref.~\cite{24}. The factor $f$ is introduced by us to adjust the theoretical level density to experiment at $E \sim B_n-1$ MeV. The parameters employed are listed in Table~1. 
From our semi-experimental partition function, the entropy can be determined from Eq.\ (\ref{eq:canentropy}). The results are shown in Fig.\ \ref{fig:fig10}. The entropy curves show a splitting at temperatures below  $k_BT = 0.5 - 0.6$ MeV which reflects the experimental splitting shown in the microcanonical plots of Figs.~7 and 8. However, the strong averaging produced by the summing in Eq.~(5), modifies the entropy excess due to components from the theoretical extrapolation of $\rho$. Even so, the curves agree qualitatively with the calculations in Fig.~5 using $\delta = 0.5$. The effect of pairing seems in both cases to vanish above $0.5 - 0.6$ MeV. This agrees with our previous work \cite{10}, giving a critical temperature of $k_BT_c =$ 0.5 MeV for the existence of pair correlations.

The observation that one quasi-particle carries 1.7 $k_B$ of entropy, can be utilized to estimate the number of quasi-particles as function of excitation energy. Analogously to Eq.~(20), we estimate from the experimental entropies $S_{\rm eo}$ and $S_{\rm ee}$ in neighboring even-odd and even-even isotopes the entropy excess $\Delta S = S_{\rm eo} - S_{\rm ee}$. The number of quasi-particles in the even-odd and even-even systems is given by Eq.\ (\ref{eq:quasi-particle}), except that the odd system is replaced by an odd-even nucleus and the even system by an even-even nucleus. 

The extracted number of quasi-particles $n(E)$ in $^{162}$Dy and $^{172}$Yb is shown in Fig.\ \ref{fig:fig11}. The number of quasi-particles raises to a level of $n \sim $ 2 around $E = 1.5 - 2$ MeV, which could be a signal for the formation of two quasi-particle states. However, the creation of four and six quasi-particles shows no clear step-like function. The breaking of additional pairs is spread out in excitation energy giving a rather smooth increase in the number of quasi-particles as function of excitation energy. In the excitation region $0.5 - 5$ MeV the $n(E)$ curve gives on the average 1.6 MeV of excitation energy to create a quasi-particle pair. This value is consistent with pairing gap parameters of this mass region, see Table 1. The theoretical calculation\footnote{The reader should keep in mind that the number of particles in the theoretical calculation and experiment are rather different. In experiment, if one assumes $^{132}$Sn as closed shell core, the number of valence protons and neutrons is of the order of $\sim 30-40$. However, performing the above theoretical calculations with say 10 or 14 particles results in qualitatively similar results as those presented here.} with $\delta =$ 0.5 gives an energy of 1.7 MeV per broken pair, which is close to the experimental finding of 1.6 MeV. Hence, with a single-particle spacing of $d = 0.1 - 0.2$ MeV, the pairing strength is determined to $G = 0.2 - 0.4$ MeV.

\section{Conclusions}

The entropy as function of excitation energy has been extracted for the $^{161,162}$Dy and $^{171,172}$Yb nuclei. The observed entropy excess in the even-odd nuclei compared to the even-even nuclei is interpreted as the entropy for a single quasi-particle (particle or hole) outside an even-even core. The entropy excess remains at a level of $\Delta S\sim$ 1.7 $k_B$ as function of excitation energy. A simple pairing model with an equidistant level spacing of $d$ and a pairing strength of $G$, gives a qualitatively similar description of these features.

The number of excited quasi-particles has been extracted from data. The onset of two quasi-particle excitations seems evident; however, the breaking of additional pairs is smeared out in excitation energy and is difficult to observe. The maximum number of excited quasi-particles is measured to $n \sim$ 6 at an excitation energy of 5.5 MeV in the $^{162}$Dy and $^{172}$Yb isotopes. 

The quasi-particle picture has been a success in describing rotational bands in cold nuclei. The present results indicate that quasi-particles also can describe certain thermodynamical properties of hot nuclei. This gives hope for realistic modeling of nuclei up to high intrinsic energy with several quasi-particles excited.

The authors are grateful to E.A.~Olsen and J.~Wikne for providing the excellent experimental conditions. We wish to acknowledge the support from the Norwegian Research Council (NFR).

\end{multicols}

\begin{table}
Table 1: Parameters used in the back-shifted Fermi gas formula for the extrapolation of the experimental level density curve.\\
\begin{tabular}{l|c|c|c|c|c}
Nucleus    & $\Delta$ (keV) & $a$ (MeV$^{-1}$)& $C_1$ (keV) & $E_1$ (keV)& $f$ \\ \hline
&&&&&\\
$^{161}$Dy &       793      & 17.46           & -1298       &    -505    &  1.400     \\
$^{162}$Dy &      1847      & 17.56           & -1296       &     551    &  1.138     \\
$^{171}$Yb &       680      & 18.40           & -1273       &    -593    &  0.376     \\
$^{172}$Yb &      1606      & 18.50           & -1271       &     335    &  0.465     \\
\end{tabular}
\end{table}


\begin{figure}
\includegraphics[totalheight=17.5cm,angle=0,bb=0 20 350 730]{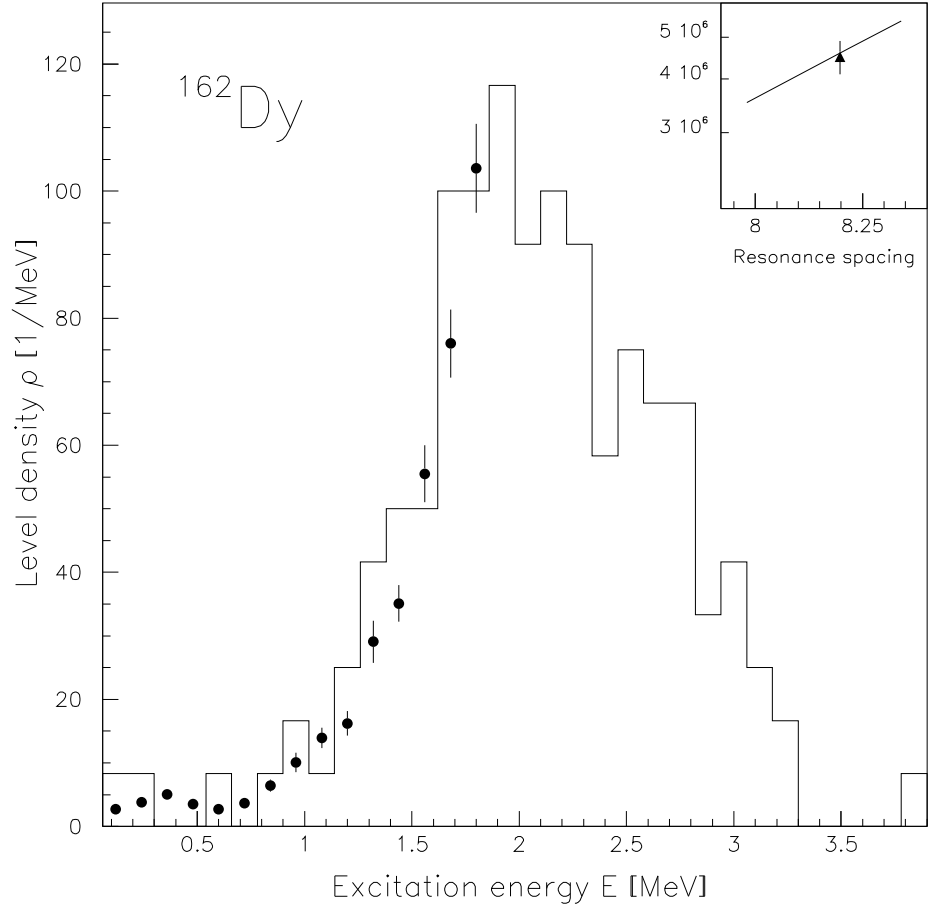}
\caption{The extracted level density for $^{162}$Dy reproduces both known levels (histogram) at low excitation energy and the neutron resonance spacing at $B_n$ (triangle) by proper adjustment of the parameters $A$ and $\alpha$ of Eq.~(2).} 
\label{fig:fig1}
\end{figure}

\begin{figure}
\includegraphics[totalheight=20cm,angle=0,bb=0 20 350 730]{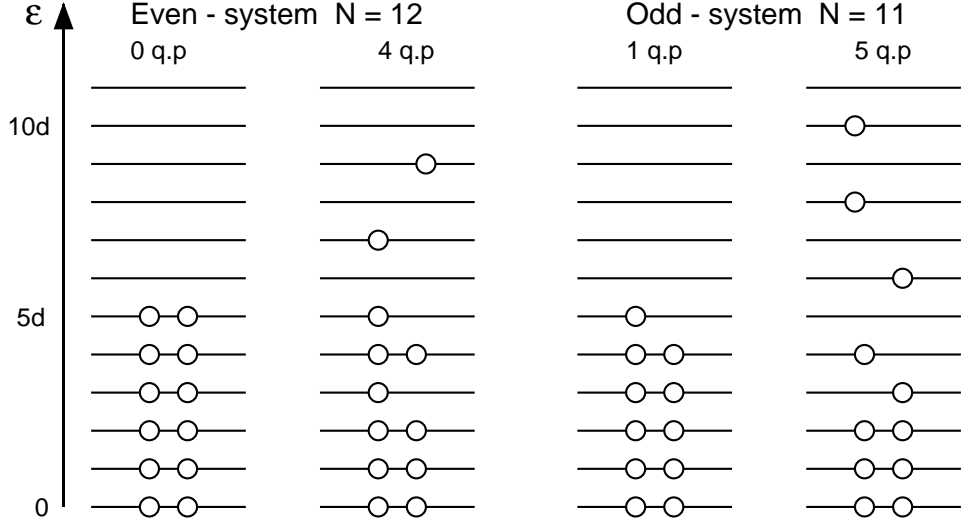}
\caption{Simple illustration of the ground state and possible excited states ( 4 and 5 quasi-particles) for a system with 12 doubly-degenerate single-particle levels. The properties of the model is governed by the level spacing $d$ and the pairing strength parameter $G$ (the illustration is with $G=0$). For the even system with 12 particles, the first excited state is a two quasi-particle state corresponding to the breaking of one pair. The first excited state in the odd system with 11 particles is a single quasi-particle state.} 
\label{fig:fig2}
\end{figure}

\begin{figure}
\includegraphics[totalheight=20cm,angle=0,bb=0 20 350 730]{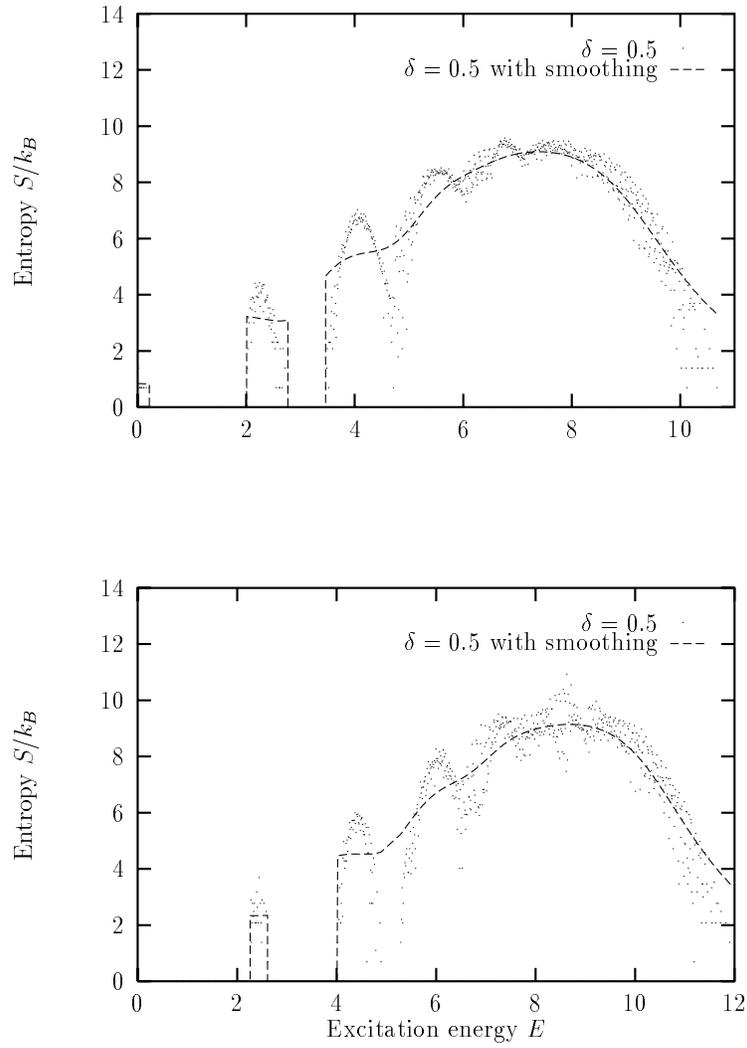}
\caption{Entropy in the microcanonical ensemble as function of excitation energy $E$ for $\delta=0.5$. The upper and lower panels show the results for the odd and even systems, respectively. Results with and without a Gaussian smoothing are displayed.} 
\label{fig:fig3}
\end{figure}

\begin{figure}
\includegraphics[totalheight=20cm,angle=0,bb=0 20 350 730]{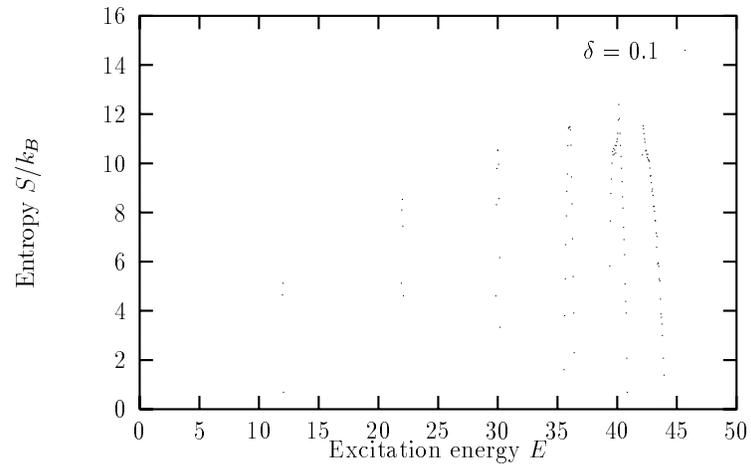}
\caption{Entropy in the microcanonical ensemble as function of excitation energy $E$ for $\delta=0.1$.} 
\label{fig:fig4}
\end{figure}

\begin{figure}
\includegraphics[totalheight=20cm,angle=0,bb=0 20 350 730]{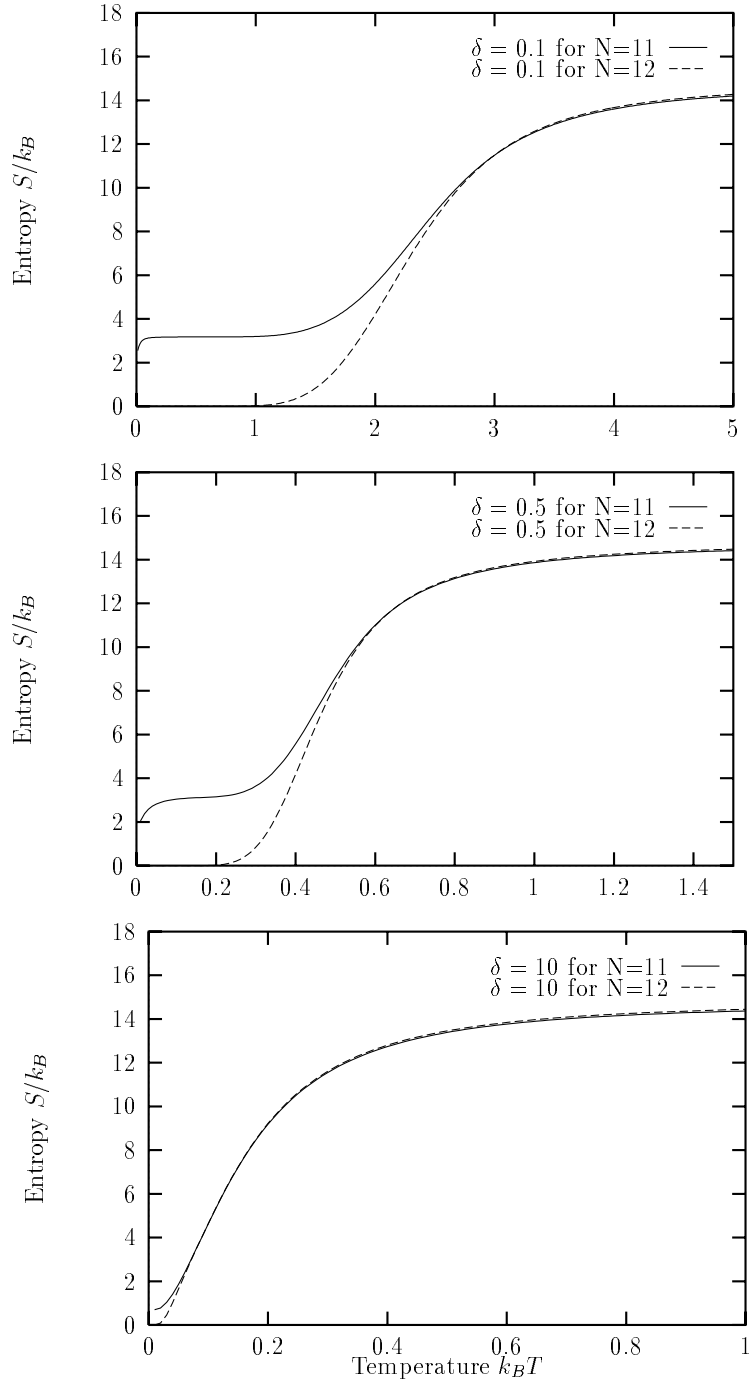}
\caption{Entropy in the canonical ensemble as function of temperature $k_BT$ for odd and even systems for $\delta=0.1$ (upper panel), $\delta=0.5$ (central panel) and $\delta=10$ (lower panel).} 
\label{fig:fig5}
\end{figure}

\begin{figure}
\includegraphics[totalheight=18cm,angle=0,bb=0 20 350 730]{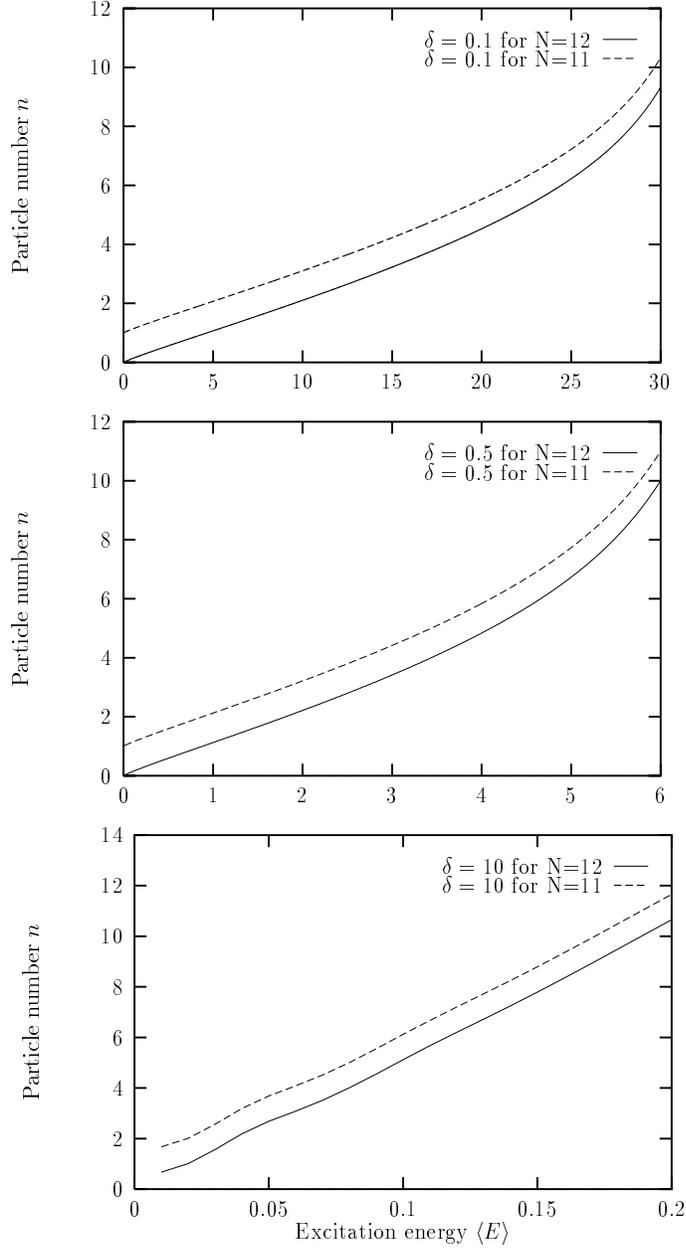}
\caption{Number of quasi-particles $n$ in the canonical ensemble for different values of $\delta$ for even and odd particle systems. Results for $\delta=0.1$ are shown in the upper panel, $\delta=0.5$ in the central panel and $\delta=10$ in the lower panel.} 
\label{fig:fig6}
\end{figure}

\begin{figure}
\includegraphics[totalheight=19cm,angle=0,bb=0 20 350 730]{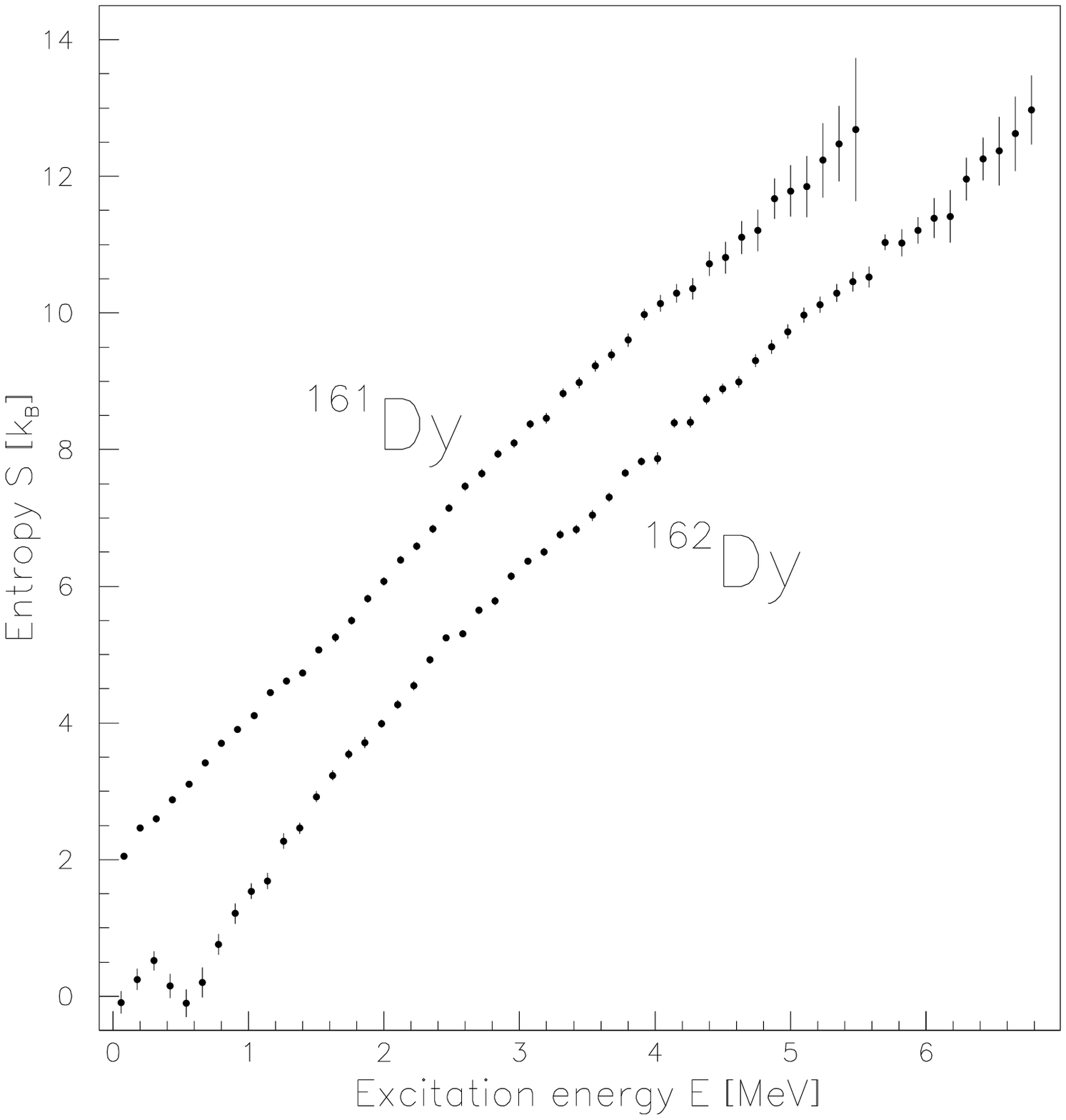}
\caption{Observed entropy for $^{161,162}$Dy as function of excitation energy $E$.}
\label{fig:fig7}
\end{figure}

\begin{figure}
\includegraphics[totalheight=19cm,angle=0,bb=0 20 350 730]{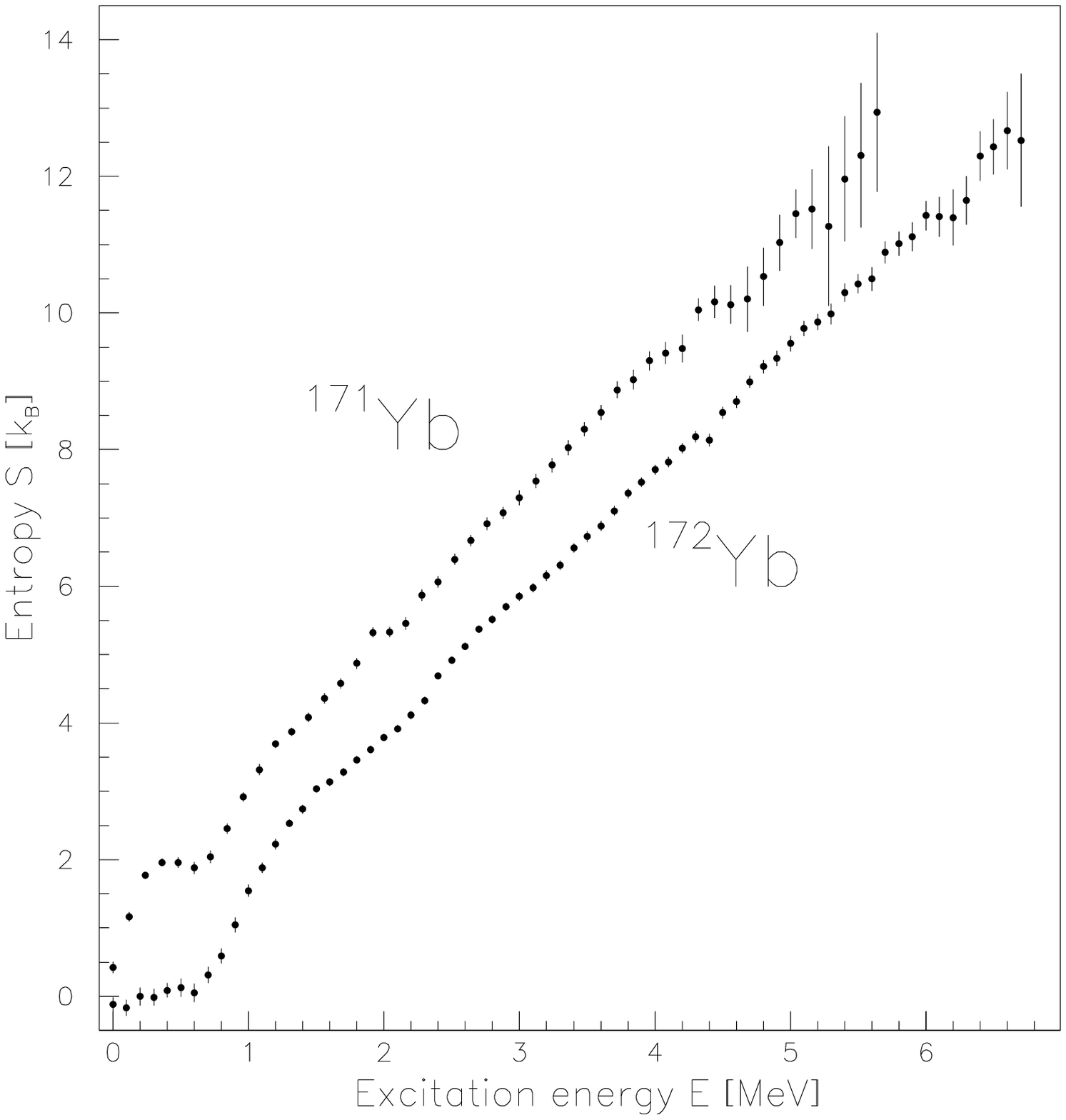}
\caption{Observed entropy for $^{171,172}$Yb as function of excitation energy $E$.}
\label{fig:fig8}
\end{figure}

\begin{figure}
\includegraphics[totalheight=17.5cm,angle=0,bb=0 80 350 730]{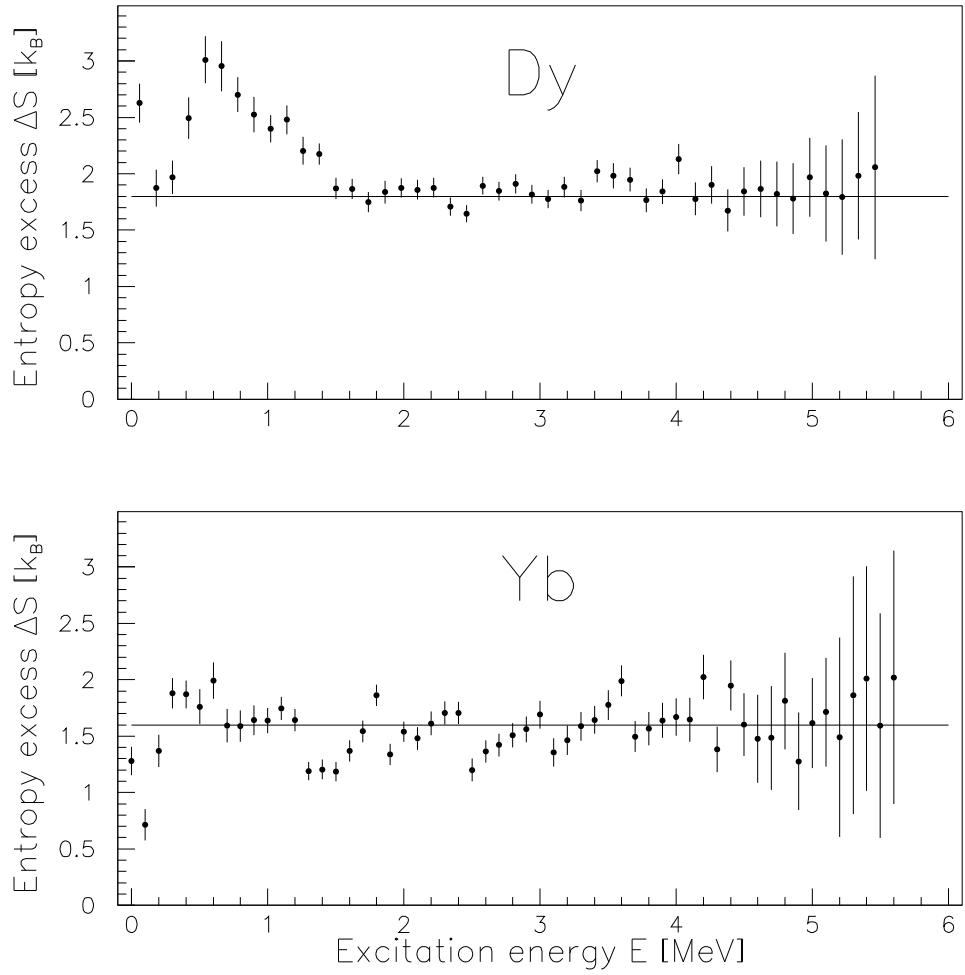}
\caption{The entropy excess $\Delta S$ in $^{161}$Dy compared to $^{162}$Dy (upper panel) and in $^{171}$Yb compared to $^{172}$Yb (lower panel). The lines through the data points indicate the average values found.}
\label{fig:fig9}
\end{figure}

\begin{figure}
\includegraphics[totalheight=17.5cm,angle=0,bb=0 80 350 730]{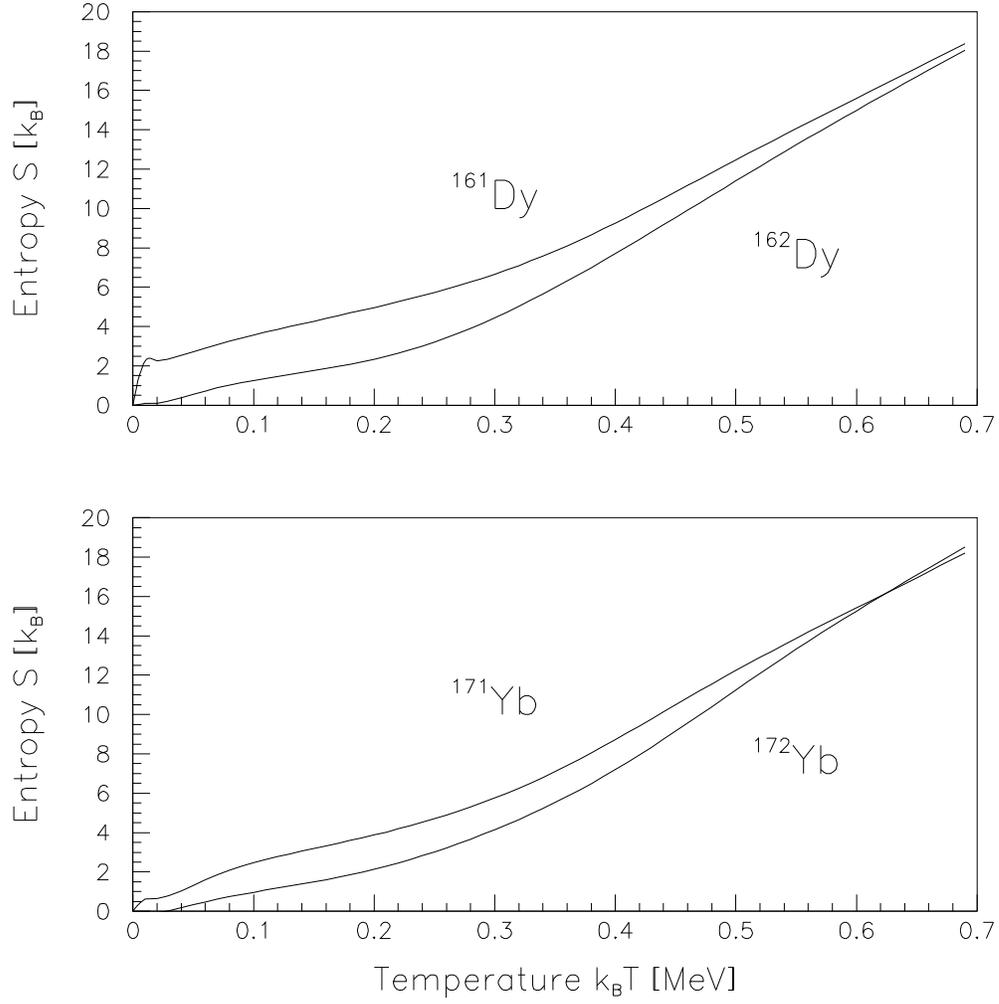}
\caption{The semi-experimental entropy $S$ for $^{161,162}$Dy and $^{171,172}$Yb calculated in the canonical ensemble as function of temperature $k_BT$.}
\label{fig:fig10}
\end{figure}

\begin{figure}
\includegraphics[totalheight=17.5cm,angle=0,bb=0 80 350 730]{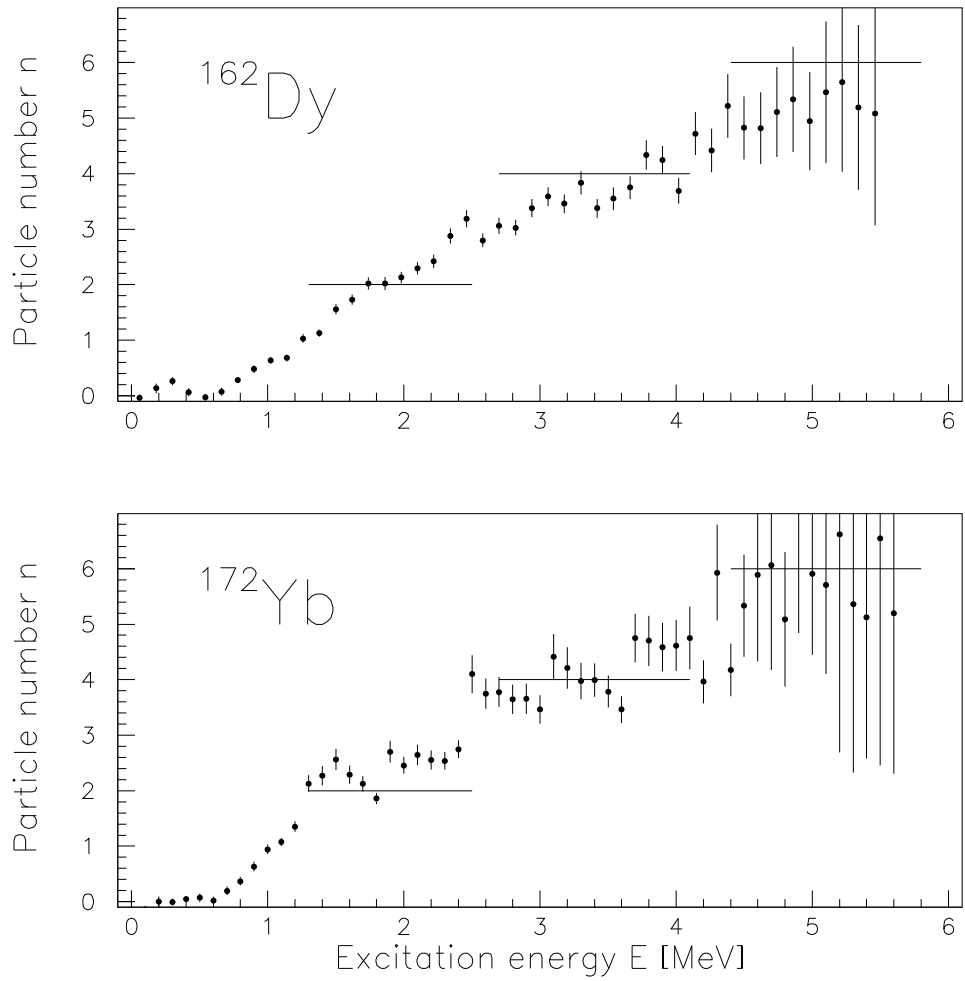}
\caption{The number of quasi-particles $n$ in $^{162}$Dy (upper panel) and $^{172}$Yb (lower panel) as function of excitation energy. The lines indicate the levels of two, four and six quasi-particles.}
\label{fig:fig11}
\end{figure}

\end{document}